\documentclass[12pt]{article}
\usepackage{amssymb,amscd,array}
\catcode `\@=11
\@addtoreset{equation}{section}

\newtheorem{thm}{Theorem}[section]

\newtheorem{prop}[thm]{Proposition}

\def\qed{\blacksquare}
\newcommand{\be}{\begin{equation}}
\newcommand{\ee}{\end{equation}}
\newcommand{\bea}{\begin{eqnarray}}
\newcommand{\eea}{\end{eqnarray}}

\newcommand{\N}{\mathbb{N}}
\newcommand{\C}{\mathbb{C}}

\begin{document}
\begin{titlepage}

\begin{center}
{\bf \Large{Multi-Graviton Theories in the Causal Approach\\}}
\end{center}
\vskip 1.0truecm
\centerline{D. R. Grigore, 
\footnote{e-mail: grigore@theory.nipne.ro}}
\vskip5mm
\centerline{Department of Theoretical Physics, Institute for Physics and Nuclear
Engineering ``Horia Hulubei"}
\centerline{Institute of Atomic Physics}
\centerline{Bucharest-M\u agurele, P. O. Box MG 6, ROM\^ANIA}

\vskip 2cm
\bigskip \nopagebreak
\begin{abstract}
\noindent
The system of multi-gravitons has been considered before in the framework of functional formalism. We consider here the
system of multi-gravitons in the causal formalism of quantum field theory. We derive in this formalism the fact that distinct gravitons 
cannot interact. The proof is based on a careful analysis of the first two orders of the perturbation theory. 
\end{abstract}
%\newpage\setcounter{page}1
\end{titlepage}

\section{Introduction}

The general framework of perturbation theory consists in the construction of 
the chronological products such that Bogoliubov axioms are verified 
\cite{BS}, \cite{EG}, \cite{DF}; for every set of Wick monomials 
$ 
A_{1}(x_{1}),\dots,A_{n}(x_{n}) 
$
acting in some Fock space
$
{\cal H}
$
one associates the operator
$$ 
T(A_{1}(x_{1}),\dots,A_{n}(x_{n})) 
$$ 
which is a distribution-valued operators called chronological product. 

The construction of the chronological products can be done recursively according
to Epstein-Glaser prescription \cite{EG}, \cite{Gl} (which reduces the induction
procedure to a distribution splitting of some distributions with causal support)
or according to Stora prescription \cite{PS} (which reduces the renormalization
procedure to the process of extension of distributions). These products are not
uniquely defined but there are some natural limitation on the arbitrariness. If
the arbitrariness does not grow with $n$ we have a renormalizable theory. A variant
based on retarded products is due to Steinmann \cite{Sto1}.

Gravity is described by particles of helicity $2$ (in the linear approximation).
Theories of higher spin are not renormalizable. However, one can save renormalizablility using ghost fields.
Such theories are defined in a Fock space
$
{\cal H}
$
with indefinite metric, generated by physical and un-physical fields (called
{\it ghost fields}). One selects the physical states assuming the existence of
an operator $Q$ called {\it gauge charge} which verifies
$
Q^{2} = 0
$
and such that the {\it physical Hilbert space} is by definition
$
{\cal H}_{\rm phys} \equiv Ker(Q)/Im(Q).
$
The space
$
{\cal H}
$
is endowed with a grading (usually called {\it ghost number}) and by
construction the gauge charge is raising the ghost number of a state. Moreover,
the space of Wick monomials in
$
{\cal H}
$
is also endowed with a grading which follows by assigning a ghost number to
every one of the free fields generating
$
{\cal H}.
$
The graded commutator
$
d_{Q}
$
of the gauge charge with any operator $A$ of fixed ghost number
\be
d_{Q}A = [Q,A]
\ee
is raising the ghost number by a unit. It means that
$
d_{Q}
$
is a co-chain operator in the space of Wick polynomials. From now on
$
[\cdot,\cdot]
$
denotes the graded commutator. From
$
Q^{2} = 0
$
one derives
\be
(d_{Q})^{2} = 0.
\label{Q-square}
\ee
 
A gauge theory assumes also that there exists a Wick polynomial of null ghost
number
$
T(x)
$
called {\it the interaction Lagrangian} such that
\be
d_{Q}T = [Q, T] = i \partial_{\mu}T^{\mu}
\label{gauge-T}
\ee
for some other Wick polynomials
$
T^{\mu}.
$
This relation means that the expression $T$ leaves invariant the physical
states, at least in the adiabatic limit. Indeed, if this is true we have:
\be
T(f)~{\cal H}_{\rm phys}~\subset~~{\cal H}_{\rm phys}  
\label{phys-inv}
\ee
up to terms which can be made as small as desired (making the test function $f$
flatter and flatter). We call this argument the {\it formal adiabatic limit}.
It is an euristic way to justify from the physical point of view relation (\ref{gauge-T}). Otherwise,
we simply have to postulate it. The preceding relation can be extended if we assume a polynomial
Poincar\'e lemma as follows. One applies 
$
d_{Q}
$
to (\ref{gauge-T}) and obtains
\bea
\partial_{\mu}d_{Q}T^{\mu} = 0
\nonumber
\eea
so we expect that we have
\bea
d_{Q}T^{\mu} = i \partial_{\nu}T^{\mu\nu}
\eea
and so on. It turns out that there are obstructions to such a polynomial Poincar\'e lemma if we work with on-shell fields, so
we must prove directly such type of identity.

One defines now the chronological products
$
T(A_{1}(x_{1}),\dots,A_{n}(x_{n})) 
$
with 
$
A_{1},\dots,A_{n}
$
of the type
$
T, T^{\mu}, T^{\mu\nu},
$
etc. and formulates a proper generalization of (\ref{gauge-T}). Such identity express, as (\ref{phys-inv}), the fact that
the scattering matrix leaves invariant the subspace of physical states, at least in some adiabatic limit sense. The analysis
of these identities can be done by direct computations in lower orders of the perturbation theory, but a general proof in
arbitrary orders is still an open problem in the general case, due to the quantum anomalies which do appear in the inductive
procedure.

Our approach is purely quantum: we do not need a classical field theory to quantize. However, we mention that there is a
variant of the perturbative quantum field theory where one keeps a closer connection with clasical theory. In this approach one
can treat quantum fields on manifolds; see for instance \cite{FR} and the recent review \cite{R}.

We will consider a system of 
$R \geq 1
$
distinct gravitions and prove that second order gauge invariance gives the following result: there is no interaction between
distinct gravitons. 
Such a result was proved for the first time in the framework of the functional formalism in \cite{BDGH}. We provide here the 
proof in the causal formalism.

\newpage

\section{Quantum Gravity\label{qg}}

The Fock space is generated by the fields
$
h^{\mu\nu}, u^{\rho}, \tilde{u}^{\sigma}
$
of null mass i.e. we have the equations of motion:
\be
\square~h^{\mu\nu} = 0 \qquad \square~u^{\rho} = 0 \qquad 
\square\tilde{u}^{\sigma} = 0
\label{KG-graviton-null-eq}
\ee
and we also assume the symmetry property
\be
h^{\mu\nu} = h^{\nu\mu}
\ee
and self-adjointness:
\be
(h^{\mu\nu})^{\dagger} = h^{\mu\nu}, \qquad 
(u^{\rho})^{\dagger} = u^{\rho}, \qquad
(\tilde{u}^{\sigma})^{\dagger} = - \tilde{u}^{\sigma}.
\label{adj-graviton-null}
\ee
We denote
\be
h \equiv \eta_{\mu\nu} h^{\mu\nu}.
\ee
The non-trivial $2$-point functions are:
\bea
<\Omega, h^{\mu\nu}(x_{1}) h^{\rho\sigma}(x_{2})\Omega> = - {i\over 2}~
(\eta^{\mu\rho}~\eta^{\nu\sigma} + \eta^{\nu\rho}~\eta^{\mu\sigma}
- \eta^{\mu\nu}~\eta^{\rho\sigma})~D_{0}^{(+)}(x_{1} - x_{2}),
\nonumber \\
<\Omega, u^{\mu}(x_{1}) \tilde{u}^{\nu}(x_{2})\Omega> = i~\eta^{\mu\nu}~
D_{0}^{(+)}(x_{1} - x_{2})
\nonumber \\
<\Omega, \tilde{u}^{\mu}(x_{1}) u^{\nu}(x_{2})\Omega> = - i~\eta^{\mu\nu}~
D_{0}^{(+)}(x_{1} - x_{2})
\label{2-graviton-null}
\eea
with
$
D_{0}(x_{1} - x_{2})
$
the Pauli-Jordan distribution and
$
D_{0}^{(\pm)}(x_{1} - x_{2})
$
the positive (and negative) frequency parts. It follows immediately:
\bea
<\Omega, h^{\mu\nu}(x_{1}) h(x_{2})\Omega> = 
<\Omega, h(x_{1}) h^{\mu\nu}(x_{2})\Omega> = i~
\eta^{\mu\nu}~D_{0}^{(+)}(x_{1} - x_{2})
\nonumber \\
<\Omega,h(x_{1}) h(x_{2})\Omega> = 4 i D_{0}^{(+)}(x_{1} - x_{2})
\eea

and the cannonical (anti)commutation relations are:
\bea
[ h^{\mu\nu}(x_{1}), h^{\rho\sigma}(x_{2}) ] = - {i\over 2}~
(\eta^{\mu\rho}~\eta^{\nu\sigma} + \eta^{\nu\rho}~\eta^{\mu\sigma}
- \eta^{\mu\nu}~\eta^{\rho\sigma})~D_{0}(x_{1} - x_{2}),
\nonumber \\
\{ u^{\mu}(x_{1}), \tilde{u}^{\nu}(x_{2}) \} = i~\eta^{\mu\nu}~D_{0}(x_{1} - x_{2})
\label{CCR-graviton-null}
\eea

We define the gauge charge according to 
\be
~[Q, h^{\mu\nu}] = - {i\over 2}~(\partial^{\mu}u^{\nu} + \partial^{\nu}u^{\mu}
- \eta^{\mu\nu} \partial_{\rho}u^{\rho}),\qquad
\{Q, u^{\mu}\} = 0,\qquad
\{Q, \tilde{u}^{\mu}\} = i~\partial_{\nu}h^{\mu\nu},
\label{Q-graviton-null}
\ee
so
\be
~[Q, h] = i~\partial_{\mu}u^{\mu}, \qquad
 ~[Q,\partial^{\nu} h_{\mu\nu}] = 0
\ee
and we can prove that the factor space
$
Ker(Q)/Im(Q)
$
describes the many-body theory of gravitions i.e. it is isomorphic to the Fock space 
$
{\cal F}(H^{[0,2]})
$
associated to the Hilbert space
$
H^{[0,2]}
$
of a particle of null mass and helicity $2$.
\newpage
We now describe the {\it off-shell} version of this construction \cite{off}, \cite{algebra}. We consider the Grassmann algebra generated by the 
variables
$
h_{\mu\nu}, u_{\mu}, \tilde{u}_{\mu}
$
with
$
h_{\mu\nu}
$
of even parity and
$
u_{\mu}, \tilde{u}_{\mu}
$
of odd parity. Next we consider the associated jet extension of order $r$
$$
h_{\mu\nu;\lambda_{1}\dots\lambda_{p}}, u_{\mu;\lambda_{1}\dots\lambda_{p}}, \tilde{u}_{\mu;\lambda_{1}\dots\lambda_{p}},
\quad p = 1,\dots,r.
$$
There is no mass constrain in this algebra. Now we define the formal derivative
$
d_{\lambda}
$
according to
\be
d_{\lambda_{0}}h_{\mu\nu;\lambda_{1}\dots\lambda_{p}} \equiv h_{\mu\nu;\lambda_{0}\dots\lambda_{p}},
\ee
etc, and define the gauge charge $Q$ by the same formula as above but with 
$
\partial_{\mu} \rightarrow d_{\mu}.
$

The operator $d_{Q}$ does not square to zero anymore. Nevertheless we define
\be
\delta T^{I} \equiv d_{\mu}T^{I\mu} 
\ee
with
$
d_{\mu}
$
the formal derivative and then
\be
s \equiv d_{Q} - i~\delta.
\ee

If we want to consider more than one specie of gravition, we only have to add a new index
$
A = 1,\dots,R
$
to the basic fields i.e. the Fock space is generated by the fields
$
h_{A}^{\mu\nu}, u_{A}^{\rho}, \tilde{u}_{A}^{\sigma}, \quad A = 1,\dots,R.
$
In the expressions for the $2$-point functions, the cannonical (anti)commutators, etc. a factor
$
\delta_{AB}
$
appears in the right-hand side i.e.
\bea
<\Omega, h_{A}^{\mu\nu}(x_{1}) h_{B}^{\rho\sigma}(x_{2})\Omega> = - {i\over 2}~\delta_{AB}
(\eta^{\mu\rho}~\eta^{\nu\sigma} + \eta^{\nu\rho}~\eta^{\mu\sigma}
- \eta^{\mu\nu}~\eta^{\rho\sigma})~D_{0}^{(+)}(x_{1} - x_{2}),
\nonumber \\
<\Omega, u_{A}^{\mu}(x_{1}) \tilde{u}_{B}^{\nu}(x_{2})\Omega> = i~\eta^{\mu\nu}~\delta_{AB}
D_{0}^{(+)}(x_{1} - x_{2})
\nonumber \\
<\Omega, \tilde{u}_{A}^{\mu}(x_{1}) u_{B}^{\nu}(x_{2})\Omega> = - i~\eta^{\mu\nu}~\delta_{AB}
D_{0}^{(+)}(x_{1} - x_{2})
\label{2-graviton-null-R}
\eea
etc.
\newpage
\section{Perturbation Theory\label{pert}}
We provide the necessary elements of (second order) of perturbation theory. Formally, 
we want to compute the scattering matrix
\be
S(g) \equiv I + i \int dx g(x) T(x) 
\nonumber \\
+ \frac{i^{2}}{ 2} \int dx~dy~g(x)~g(y)~T(x,y) + \cdots 
\label{S-matrix}
\ee
where $g$ is some test function. The expressions 
$T(x,y)$
are called {\it (second order) chronological products} because they must verify the causality property:
\be
T(x,y) = T(x) T(y)
\ee
for 
$x \succ y$
i.e. 
$(x - y)^{2} \geq 0, x^{0} - y^{0} \geq 0$;
in other words the point $x$ succeeds causally the point $y$. This is a relativistic generalization of the property
\be
U(t,s) = U(t,r) U(r,s),~~t > r > s
\ee
of the time evolution operator from non-relativistic quantum mechanics.

We go to the second order of perturbation theory using the {\it causal commutator}
\be
D^{A,B}(x,y) \equiv D(A(x),B(y)) = [ A(x),B(y)]
\ee
where 
$
A(x), B(y)
$
are arbitrary Wick monomials. These type of distributions are translation invariant i.e. they depend only on 
$
x - y
$
and the support is inside the light cones:
\be
supp(D) \subset V^{+} \cup V^{-}.
\ee

A theorem from distribution theory guarantees that one can causally split this distribution:
\be
D(A(x),B(y)) = A(A(x),B(y)) - R(A(x),B(y)).
\ee
where:
\be
supp(A) \subset V^{+} \qquad supp(R) \subset V^{-}.
\ee
The expressions 
$
A(A(x),B(y)), R(A(x),B(y))
$
are called {\it advanced} resp. {\it retarded} products. They are not uniquely defined: 
one can modify them with {\it quasi-local terms} i.e. terms proportional with
$
\delta(x - y)
$
and derivatives of it. 

There are some limitations on these redefinitions coming from Lorentz invariance and {\it power counting}: 
this means that we should not make the various distributions appearing in the advanced and retarded products too singular.

Then we define the {\it chronological product} by:
\be
T(A(x),B(y)) = A(A(x),B(y)) + B(y) A(x) = R(A(x),B(y)) + A(x) B(y).
\ee

The expression
$
T(x, y)
$
corresponds to the choice
\be
T(x,y) \equiv T(T(x), T(x)).
\ee

The ``naive'' definition
\be
T(A(x),B(y)) = \theta(x^{0}-y^{0}) A(x) B(y) + \theta(y^{0}-x^{0}) B(y) A(x)
\ee
involves an illegal operation, namely the multiplication of distributions. 
This appears in some loop contributions (the famous ultraviolet divergences). 

We will need in the following the causal commutator
\be
D^{IJ}(x,y) \equiv D(T^{I}(x), T^{J}(y)) = [T^{I}(x), T^{J}(y)]
\label{causal-comm-2}
\ee
where $[\cdot,\cdot]$ is always the graded commutator.

The chronological products
$
T(A(x),B(y))
$
must satisfy some axioms (Bogoliubov):

\begin{itemize}
\item
{\bf The ``initial condition"}:
\be
T(A(x)) = A(x).
\ee
\item
{\bf Skew-symmetry} in all arguments:
\be
T(A(x),B(y)) = (-1)^{f(A)f(B)} T(B(y),A(x))
\ee
where
$
f(A)
$
is the Fermi number of the Wick monomial $A$.
\item
{\bf Poincar\'e invariance}: we have a natural action of the Poincar\'e group in the
space of Wick monomials and we impose that for all 
$g \in inSL(2,\C)$
we have:
\be
U_{g} T(A(x),B(y)) U^{-1}_{g} = T(g\cdot A(x),g\cdot B(y))
\label{invariance}
\ee
where in the right hand side we have the natural action of the Poincar\'e group on Wick monomials.

\item
{\bf Causality}: if 
$
x - y 
$
is in the upper causal cone then we denote this relation by
$
x \succeq y
$.
In this case we have the factorization property:
\be
T(A(x),B(y)) = A(x) B(y)
\ee

\item
{\bf Unitarity}: We define the {\it anti-chronological products} 
is an ordered subset, we define
\be
\bar{T}(A(x),B(y)) \equiv A(x) B(y) + (-1)^{f(A)f(B)} B(y) A(x) - T(A(x),B(y)) 
\ee
Then the unitarity axiom is:
\be
\bar{T} = T^{\dagger}.
\label{unitarity}
\ee

\item
{\bf Power counting}: We can also include in the induction hypothesis a limitation on the order of
singularity of the vacuum averages of the chronological products associated to
arbitrary Wick monomials; explicitly:
\be
\omega(<\Omega, T(A(x),B(y))\Omega>) \leq \omega(A) + \omega(B) - 4
\label{power}
\ee
where by
$\omega(d)$
we mean the order of singularity of the (numerical) distribution $d$ and by
$\omega(A)$
we mean the canonical dimension of the Wick monomial $W$.

\item
{\bf Wick expansion property}: we refer to the literature for the formulation. 
\end{itemize}
The axioms can be extended to arbitrary chronological products
$
T(A_{1}(x_{1}),\dots,A_{n}(x_{n}).
$
Now we can construct the chronological products
$
T(T^{I_{1}}(x_{1}),\dots,T^{I_{n}}(x_{n}))
$
according to the recursive procedure. 

We say that the theory is gauge invariant
in all orders of the perturbation theory if the following set of identities:
\be
d_{Q}T(T^{I_{1}}(x_{1}),\dots,T^{I_{n}}(x_{n})) = 
i \sum_{l=1}^{n} (-1)^{s_{l}} {\partial\over \partial x^{\mu}_{l}}
T(T^{I_{1}}(x_{1}),\dots,T^{I_{l}\mu}(x_{l}),\dots,T^{I_{n}}(x_{n}))
\label{gauge-n}
\ee
are true for all 
$n \in \N$
and all
$
I_{1}, \dots, I_{n}.
$
Here we have defined
\be
s_{l} \equiv \sum_{j=1}^{l-1} f(T^{I_{j}}) = \sum_{j=1}^{l-1} |I_{j}|.
\ee
In particular, the case
$
I_{1} = \dots = I_{n} = \emptyset
$
it is sufficient for the gauge invariance of the scattering matrix, at least
in the adiabatic limit: we have the same argument as for relation (\ref{phys-inv}).

To describe this property in a cohomological framework, we consider that the chronological products are the 
cochains and we define for the operator $\delta$ by 
\be
\delta T(T^{I_{1}}(x_{1}),\dots,T^{I_{n}}(x_{n})) = 
i \sum_{l=1}^{n} (-1)^{s_{l}} {\partial\over \partial x^{\mu}_{l}}
T(T^{I_{1}}(x_{1}),\dots,T^{I_{l}\mu}(x_{l}),\dots,T^{I_{n}}(x_{n})).
\label{der}
\ee
It is easy to prove that we have:
\be
\delta^{2} = 0
\ee
and
\be
[ d_{Q}, \delta ] = 0.
\ee
Next we define 
\be
s \equiv d_{Q} - i \delta
\ee
such that relation (\ref{gauge-n}) can be rewritten as
\be
sT(T^{I_{1}}(x_{1}),\dots,T^{I_{n}}(x_{n})) = 0.
\label{s-gauge-n}
\ee

We note that if we define
\be
\bar{s} \equiv d_{Q} + i \delta
\ee
we have
\be
s\bar{s} = 0, \qquad \bar{s} s = 0
\label{s2}
\ee
so expressions verifying the relation
$
s C = 0
$
can be called {\it cocycles} and expressions of the type
$
\bar{s}B
$
are the {\it coboundaries}. One can build the corresponding cohomology space in the standard way.

If we have (\ref{s-gauge-n}) for 
$
n = 1,2,\dots, p - 1
$
then the relation (\ref{s-gauge-n}) for 
$
n = p
$ can be broken by anomalies i.e.we have:
\be
sT(T^{I_{1}}(x_{1}),\dots,T^{I_{p}}(x_{p})) = {\cal A}^{I_{1},\dots,I_{p}}(x_{1},\dots,x_{p})
\label{s-gauge-n-ano}
\ee
where 
$
{\cal A}^{I_{1},\dots,I_{p}}
$
is a quasi-local expression, having support in
\be
D_{p} = \{ x_{1} = x_{2} = \dots = x_{p} \}.
\ee

The gauge theory is physically meaningful if one can remove the anomalies by a redefinition of the chronological products.
\newpage
\section{First-Order\label{first}}

\begin{prop}
Suppose that $T$ is a Wick polynomial in the (quantum) fields
$
h_{A}^{\mu\nu}, u_{A}^{\rho}, \tilde{u}_{A}^{\sigma}
$
($
A = 1,\dots,R.
$)
which is: (a) Poincar\'e invariant; (b) of canonical dimension
$
\omega(T) \leq 5;
$
(c) trilinear in the fields; (d) self-adjoint; (e) gauge invariant in the sense (\ref{gauge-T}). Then one can take
\bea
T = t^{ABC}~(h_{A}^{\mu\nu} \partial_{\mu}h_{B} \partial_{\nu}h_{C}
- 2 h^{\mu\nu}_{A} \partial_{\mu}h_{B\rho\sigma} \partial_{\nu}h_{C}^{\rho\sigma}
- 4 h_{A\mu\nu} \partial_{\rho}h_{B}^{\mu\sigma} \partial_{\sigma}h_{C}^{\nu\rho}
\nonumber \\
+ 4 \partial_{\mu}h_{A}^{\mu\nu} u_{B}^{\rho} \partial_{\rho}\tilde{u}_{C\nu}
- 4 h_{A}^{\mu\nu} \partial_{\mu}u_{B}^{\rho} \partial_{\nu}\tilde{u}_{C\rho} )
\label{first-order-a}
\eea
up to a coboundary. Here
$
t^{ABC},~A, B, C = 1,\dots,R
$
are some real constants with complete symmetry in the indexes.
\label{Ta}
\end{prop}
{\bf Proof:}
It follows easily that we can have terms with 
$
\omega = 3, 5.
$

(i) In the case 
$
\omega = 3
$
we can have the following terms:
\bea
T_{1} = t^{ABC}_{1}~h^{\mu\nu}_{A} h_{B\nu\rho} h_{C\mu}^{\rho}, \quad 
\nonumber\\
T_{2} = t^{ABC}_{2}~h^{\mu\nu}_{A} h_{B\mu\nu} h_{C},
\nonumber\\
T_{3} = t^{ABC}_{3}~h_{A} h_{B} h_{C}, 
\nonumber\\
T_{4} = t^{ABC}_{4}~h_{A}^{\mu\nu} u_{B\mu} \tilde{u}_{C\nu}
\nonumber\\
T_{5} = t^{ABC}_{5}~h_{A} u_{B\mu} \tilde{u}_{C}^{\mu}
\eea
where
\bea
t^{ABC}_{j} = A \leftrightarrow B = B \leftrightarrow C, \quad j = 1, 3.
\nonumber\\
\quad t^{ABC}_{2} = A \leftrightarrow B.
\eea

By simple ``partial integration" we can write 
\be
d_{Q}T = i \partial_{\mu}X^{\mu} + i u_{A}^{\mu} Y_{\mu}^{A} + i Z
\label{dqt}
\ee
with easily computable expresions
$
Y \sim h h, Z \sim u u \tilde{u}.
$
The equation (\ref{gauge-T}) becomes
\be
u_{A}^{\mu} Y_{\mu}^{A} + Z = \partial_{\mu}\tilde{T}^{\mu}, \quad \tilde{T}^{\mu} \equiv T^{\mu} - X^{\mu}.
\label{Y3}
\ee
The generic form for 
$
\tilde{T}^{\mu} 
$
is:
\be
\tilde{T}^{\mu} = u_{A\nu} T^{\mu\nu}_{A} + S^{\mu}
\ee
with
$
T^{\mu\nu}_{A} \sim h h, S^{\mu} \sim u u \tilde{u}.
$

If we introduce in (\ref{Y3}) then we easily obtain
$
T^{\mu\nu}_{A} = 0
$
and from here
$
Y_{\mu}^{A} = 0.
$
If we write explicitly this equation then it immediately follows that
$
t_{j} = 0, \quad j = 1,\dots,5
$
so there is no solution with canonical dimension $3$.
\newpage
(ii) For 
$
\omega = 5
$
we have terms of the type 
$
h h h
$ 
and 
$
h u \tilde{u}
$
with two derivatives distributed on the three factors. We can simplify the list using the following observations (see \cite{massive}).

- We can consider for the first type only terms of the form
$
h \partial h \partial h
$
because we can eliminate the terms of the form
$
h h \partial\partial h
$
subtracting a total derivative. In the same way we can consider only terms of the type
$
u \partial\tilde{u} \partial h, u \partial\partial\tilde{u} h, u \tilde{u} \partial\partial h.
$

- We can use the identity:
\bea
\square f_{j} = 0, \quad j = 1,2,3 
\quad \Longrightarrow
%\nonumber \\
(\partial^{\mu}f_{1}) (\partial_{\mu}f_{2}) f_{3}
= {1\over 2} \partial_{\mu} \Bigl[ (\partial^{\mu}f_{1}) f_{2} f_{3}
+ f_{1} (\partial^{\mu}f_{2}) f_{3} - 
f_{1} f_{2} (\partial^{\mu}f_{3}) \Bigl]
\label{magic}
\eea
to eliminate some terms.

- We can list all possible expressions of the type
$
d_{Q}L
$
with
$
\omega(L) = 4
$
and
$
gh(L) = - 1
$
to eliminate another set of terms. We are left with the following list: 
\bea
T_{1} = t^{ABC}_{1}~h^{\mu\nu}_{A} \partial_{\mu}h_{B\rho\sigma} \partial_{\nu}h_{C}^{\rho\sigma} \qquad
%\nonumber \\
T_{2} = t^{ABC}_{2}~h_{A\mu\nu} \partial_{\rho}h_{B}^{\mu\sigma} \partial_{\sigma}h_{C}^{\nu\rho}
\nonumber \\
T_{3} = t^{ABC}_{3}~h_{A} \partial^{\mu}h_{B}^{\rho\sigma} \partial_{\rho}h_{C\mu\sigma} \qquad
%\nonumber \\
T_{4} = t^{ABC}_{4}~h_{A}^{\mu\nu} \partial^{\rho}h_{B} \partial_{\mu}h_{C\nu\rho}
\nonumber \\
T_{5} = t^{ABC}_{5}~h_{A}^{\mu\nu} \partial_{\mu}h_{B} \partial_{\nu}h_{C} \qquad
%\nonumber \\
T_{6} = t^{ABC}_{6}~u_{A}^{\mu} \partial^{\nu}h_{B\mu\nu} \partial_{\rho}\tilde{u}_{C}^{\rho}
\nonumber\\
T_{7} = t^{ABC}_{7}~u_{A}^{\mu} \partial_{\rho}h_{B}^{\nu\rho} \partial_{\nu}\tilde{u}_{C\mu} \qquad
%\nonumber\\
T_{8} = t^{ABC}_{8}~u_{A}^{\mu} \partial_{\rho}h_{B}^{\nu\rho} \partial_{\mu}\tilde{u}_{C\nu}
\nonumber\\
T_{9} = t^{ABC}_{9}~u_{A}^{\mu} \partial_{\rho}h_{B\mu\sigma} \partial^{\sigma}\tilde{u}_{C}^{\rho} \qquad
%\nonumber\\
T_{10} = t^{ABC}_{10}~u_{A}^{\mu} \partial_{\mu}h_{B\rho\sigma} \partial^{\rho}\tilde{u}_{C}^{\sigma}
\nonumber\\
T_{11} = t^{ABC}_{11}~u_{A}^{\mu} h_{B\mu\nu} \partial^{\nu}\partial_{\rho}\tilde{u}_{C}^{\rho} \qquad
%\nonumber\\
T_{12} = t^{ABC}_{12}~u_{A}^{\mu} h_{B}^{\rho\sigma} \partial_{\mu}\partial_{\rho}\tilde{u}_{C\sigma}
\nonumber\\
T_{13} = t^{ABC}_{13}~u_{A}^{\mu} h_{B}^{\rho\sigma} \partial_{\rho}\partial_{\sigma}\tilde{u}_{C\mu} \qquad
%\nonumber\\
T_{14} = t^{ABC}_{14}~u_{A}^{\mu} \partial_{\mu}h_{B} \partial_{\nu}\tilde{u}_{C}^{\nu}
\nonumber\\
T_{15} = t^{ABC}_{15}~u_{A}^{\mu} \partial_{\nu}h_{B} \partial_{\mu}\tilde{u}_{C}^{\nu} \qquad
%\nonumber\\
T_{16} = t^{ABC}_{16}~u_{A}^{\mu} h_{B} \partial_{\mu}\partial_{\nu}\tilde{u}_{C}^{\nu}
\nonumber\\
T_{17} = t^{ABC}_{17}~u_{A}^{\mu} \partial_{\mu}\partial_{\nu}h_{B} \tilde{u}_{C}^{\nu}
\eea
\be
t^{ABC}_{j} = B \leftrightarrow C, \qquad j = 1, 2, 3, 5.
\ee

We can proceed as above and obtain (\ref{dqt}) and (\ref{Y3}), but now the generic form of
$
\tilde{T}^{\mu}
$
is more complicated:
\be
\tilde{T}^{\mu} = u_{A\nu}~T_{A}^{\mu\nu} + (\partial_{\rho}u_{A\nu})~T_{A}^{\mu\nu\rho} 
+ (\partial_{\rho}\partial_{\sigma}u_{A\nu})~T_{A}^{\mu\nu\rho\sigma}
+ S^{\mu} 
\label{tmu}
\ee
where 
$
T_{A}^{\mu\nu} , T_{A}^{\mu\nu\rho}, T_{A}^{\mu\nu\rho\sigma} 
$
are bilinear in $h$ and 
$
S^{\mu} \sim u u \tilde{u}.
$
Moreover we can assume that
$
T_{A}^{\mu\nu\rho\sigma} = \rho \leftrightarrow \sigma.
$
%\newpage
By direct computations we get from (\ref{Y3})
\bea
Y_{A}^{\nu} = \partial_{\mu}T_{A}^{\mu\nu}
\nonumber \\
T_{A}^{\rho\nu} = - \partial_{\mu}T_{A}^{\mu\nu\rho}
\nonumber\\
\partial_{\rho}\partial_{\sigma}u_{A\nu}~(T_{A}^{\sigma\nu\rho} + \partial_{\mu}T_{A}^{\mu\nu\rho\sigma}) = 0
\nonumber\\
\partial_{\mu}\partial_{\rho}\partial_{\sigma}u_{A\nu}~T_{A}^{\mu\nu\rho\sigma} = 0.
\label{Y3a}
\eea
From the first two equations we obtain
\be
Y_{A}^{\mu} = - \partial_{\rho}\partial_{\sigma}T_{A}^{\rho\mu\sigma}
\label{Ya}
\ee
so we can suppose that
$
T_{A}^{\rho\mu\sigma} = \rho \leftrightarrow \sigma.
$
We write
\bea
T_{A}^{\rho\mu\sigma} = t^{\rho\mu\sigma} + \eta^{\rho\sigma}~t_{A}^{\mu}
\nonumber\\
T_{A}^{\mu\nu\rho\sigma} = t_{A}^{\mu\nu\rho\sigma} + \eta^{\rho\sigma}~t_{A}^{\mu\nu}
\eea
where the expressions
$
t^{\rho\mu\sigma}, t_{A}^{\mu\nu\rho\sigma}
$
do not contain terms with the factor
$
\eta^{\rho\sigma}
$
and we have the symmetry properties
$
t^{\rho\mu\sigma} = \rho \leftrightarrow \sigma,\quad t_{A}^{\mu\nu\rho\sigma} = \rho \leftrightarrow \sigma.
$
From (\ref{Ya}) we obtain
\be
Y_{A}^{\mu} = - \partial_{\rho}\partial_{\sigma}t_{A}^{\rho\mu\sigma} - \square t_{A}^{\mu}
\label{Yb}
\ee
and from the third equation of the system (\ref{Y3a}) it follows
\be
t_{A}^{\rho\nu\sigma} = - \partial_{\mu}t_{A}^{\mu\nu\rho\sigma}
\ee
so the previous relation becomes:
\be
Y_{A}^{\mu} = \partial_{\nu}\partial_{\rho}\partial_{\sigma}t_{A}^{\nu\mu\rho\sigma} - \square t_{A}^{\mu}.
\label{Yc}
\ee
But the last relation of the system (\ref{Y3a}) shows that the first term in the right hand side is null, so we have:
\be
Y_{A}^{\mu} = - \square t_{A}^{\mu}.
\label{Yd}
\ee
Because 
$
t_{A}^{\mu}
$
is bilinear in $h$ the expression
$
\square t_{A}^{\mu}
$
will be a sum of terms of the type
$
\partial_{\mu}f\partial^{\mu}g.
$
So we have to determine the expression
$
Y_{A}^{\mu}
$
by direct computation: we compute the expressions
$
d_{Q}T_{j}
$
and by ``partial integration" put them in the standard form from (\ref{dqt}). Then we must consider only the terms which are {\bf not} 
of the form
$
\partial_{\mu}f\partial^{\mu}g,
$
group them and put the result to zero.

The solution of the system is the following: 
\bea
t^{ABC}_{1} = - 2~t^{ABC},\quad t^{ABC}_{2} = - 4~t^{ABC},\quad t^{ABC}_{5} = t^{ABC},
\nonumber\\
t^{ABC}_{7} = 4~t^{ABC} \quad t^{ABC}_{8} = 4~t^{ABC} \quad t^{ABC}_{13} = 4~t^{ABC}
\eea
with 
$
t^{ABC}
$
having the property of complete symmetry. We can rewrite
$
T_{7} + T_{13}
$
as the fourth term from the expresion $T$ from the statement plus a total derivative which we can eliminate.
$\qed$
%\newpage

The expresion $T$ cannot be easily compared to the classical expresion derived from Hilbert-Einstein Lagrangian. As remarked in
\cite{massive} we can add two terms of the type (\ref{magic}); also we can rewrite the ghost contribution, all amounting to the  elimination
of a total derivative. In this case we get the expresion from \cite{descent}:
\begin{prop}
\bea
T = t^{ABC}~( h_{A}^{\mu\nu} \partial_{\mu}h_{B} \partial_{\nu}h_{C}
- 2 h^{\mu\nu}_{A} \partial_{\mu}h_{B\rho\sigma} \partial_{\nu}h_{C}^{\rho\sigma}
- 4 h_{A\mu\nu} \partial_{\rho}h_{B}^{\mu\sigma} \partial_{\sigma}h_{C}^{\nu\rho}
\nonumber \\
- 2 h_{A}^{\mu\nu} \partial_{\rho}h_{B\mu\nu} \partial^{\rho}h_{C}
+ 4 h_{A}^{\mu\nu} \partial_{\sigma}h_{B\mu\rho} \partial^{\sigma}h_{C\nu}^{\rho}
\nonumber\\
- 4 h_{A}^{\mu\nu} \partial_{\mu}u_{B}^{\rho} \partial_{\rho}\tilde{u}_{C\nu}
+ 4 \partial^{\rho}h_{A}^{\mu\nu} u_{B\rho} \partial_{\mu}\tilde{u}_{C\nu} 
+ 4 h_{A}^{\mu\nu} \partial_{\rho}u_{B}^{\rho} \partial_{\mu}\tilde{u}_{C\nu}
- 4 h_{A}^{\mu\nu} \partial_{\mu}u_{B}^{\rho} \partial_{\nu}\tilde{u}_{C\rho}).
\label{first-order-b}
\eea

\bea
T^{\mu} = t^{ABC}~[ u_{A}^{\nu} ( 2 \partial^{\mu}h_{B}^{\rho\sigma} \partial_{\nu}h_{C\rho\sigma}
- 4 \partial_{\nu}h_{B\rho\sigma} \partial^{\rho}h_{C}^{\mu\sigma}
-  \partial^{\mu}h_{B} \partial_{\nu}h_{C} )
\nonumber \\
+ u_{A}^{\mu} \left( {1\over 2} \partial_{\nu}h_{B} \partial^{\nu}h_{C}
- \partial_{\nu}h_{B\rho\sigma} \partial^{\nu}h_{C}^{\rho\sigma}
+ 2 \partial_{\rho}h_{B\nu\sigma} \partial^{\sigma}h_{C}^{\nu\rho}
 \right)
\nonumber \\
+ \partial_{\rho}u_{A\nu} ( 4 h_{B}^{\rho\sigma} \partial_{\sigma}h_{C}^{\mu\nu}
- 4 h_{B}^{\rho\sigma} \partial^{\mu}h_{C\sigma}^{\nu}
+ 2 h_{B}^{\nu\rho} \partial^{\mu}h_{C} 
+ 4 h_{B}^{\rho\sigma} \partial^{\nu}h^{\mu}_{C\sigma} )
\nonumber \\
+ \partial_{\nu}u_{A}^{\nu} ( 2 h_{B\rho\sigma} \partial^{\mu}h_{C}^{\rho\sigma}
 - h_{B} \partial^{\mu}h_{C}
- 4 \partial^{\rho}h_{B}^{\mu\sigma} h_{C\rho\sigma}) 
\nonumber \\
+ 2 u_{A}^{\nu} ( \partial_{\nu}u_{B\rho} \partial^{\mu}\tilde{u}_{C}^{\rho}
-  \partial^{\rho}u_{B}^{\mu} \partial_{\nu}\tilde{u}_{C\rho}
- \partial_{\nu}\partial_{\rho}u_{B}^{\rho} \tilde{u}_{C}^{\mu})
\nonumber\\
+ 2 u_{A}^{\mu} ( \partial_{\nu}u_{B\rho} \partial^{\rho}\tilde{u}_{C}^{\nu}
+  \partial_{\nu}\partial^{\rho}u_{B}^{\mu} \tilde{u}_{C}^{\nu} )
\nonumber\\
- 2 \partial_{\nu}u_{A}^{\nu} \partial^{\rho}u_{B}^{\mu} \tilde{u}_{C\rho} ].
\eea
\bea
T^{\mu\nu} = 2~t^{ABC}~\{[ ( - u_{A}^{\rho} \partial_{\rho}u_{B\sigma} \partial^{\nu}h_{C}^{\mu\sigma}
- u_{A}^{\rho} \partial_{\sigma}u_{B}^{\mu} \partial_{\rho}h_{C}^{\nu\sigma}
+ u_{A}^{\mu} \partial_{\rho}u_{B\sigma} \partial^{\sigma}h_{C}^{\nu\rho}
\nonumber\\
- \partial_{\rho}u_{A}^{\rho} \partial_{\sigma}u_{B}^{\mu} h_{C}^{\nu\sigma} ) - (\mu \leftrightarrow \nu) ]\}
- 4 \partial^{\rho}u_{A}^{\mu} \partial^{\sigma}u_{B}^{\nu} h_{C\rho\sigma}
\eea
\bea
T^{\mu\nu\rho} = t^{ABC}~[ u_{A\lambda} ( \partial^{\mu}u_{B}^{\nu} \partial^{\lambda}u_{C}^{\rho}
+ \partial^{\nu}u_{B}^{\rho} \partial^{\lambda}u_{C}^{\mu}
+ \partial^{\rho}u_{B}^{\mu} \partial^{\lambda}u_{C}^{\nu} )
\nonumber\\
- \partial^{\nu}u_{A}^{\lambda} \partial_{\lambda}u_{B}^{\mu} u_{C}^{\rho}
- \partial^{\rho}u_{A}^{\lambda} \partial_{\lambda}u_{B}^{\nu} u_{C}^{\mu}
- \partial^{\mu}u_{A}^{\lambda} \partial_{\lambda}u_{B}^{\rho} u_{C}^{\nu}) - (\mu \leftrightarrow \nu) ]
\eea
\label{T}
\end{prop}
In the off-shell formalism \cite{off} we have the following expressions for 
\be
S^{I} \equiv sT^{I} = d_{Q}T^{I} - i \partial_{\mu}T^{I\mu}.
\ee
\bea
S = i t^{ABC} ( - 2 u_{A}^{\mu}~d_{\mu}h_{B\alpha\beta}~\square h_{C}^{\alpha\beta}
- 2 d_{\mu}u_{A}^{\mu}~h_{B\alpha\beta}~\square h_{C}^{\alpha\beta}
+ u_{A}^{\mu}~d_{\mu}h_{B}~\square h_{C}
+ d_{\mu}u_{A}^{\mu}~h_{B}~\square h_{C}
\nonumber\\
- 2 d_{\alpha}u_{A\beta}~h_{B}^{\alpha\beta}~\square h_{C}
+ 4 d^{\rho}u_{A\mu}~h_{B\nu\rho}~\square h_{C}^{\mu\nu}
- 2 u_{A}^{\mu}~d_{\mu}u_{B\nu}~\square\tilde{u}_{C}^{\nu}
\eea
\be
S^{\mu} \equiv 2 i~u_{A}^{\rho}~d_{\rho}u_{B\nu}~\square h_{C}^{\mu\nu}
\ee
\be
S^{[\mu\nu]} = i~(u_{A}^{\rho}~d_{\rho}u_{B}^{\nu} \square u_{C}^{\mu}
- \square u_{A}^{\mu} \partial_{\rho}u_{B}^{\nu}~\square u_{C}^{\rho}) - (\mu \leftrightarrow \nu)
\ee
Here 
$
\square
$
is the formal d'Alembert operator build from formal derivatives:
\be
\square \equiv d_{\mu}d^{\mu}.
\ee
\newpage

\section{Second Order\label{second}}
We compute the gauge variation in the off-shell formalism. This means that we compute off-shell the expresion
$
sT(T(x),T(y)).
$
The result is:
\be
sT(T(x),T(y)) = \square D^{F}(x - y) [ A(x,y) + (x \leftrightarrow y) ] 
+ \partial_{\mu}\square D^{F}(x - y) [ A^{\mu}(x,y) - (x \leftrightarrow y) ]  
\ee
where
\be
A = A_{h} + A_{gh}, \qquad A^{\mu} = A_{h}^{\mu} + A_{gh}^{\mu}
\ee
The explicit expressions are:
\bea
A_{h}(x,y) = 2~t^{ABCD}~[ (u_{A}^{\lambda}\partial_{\lambda}h_{B}^{\mu\nu})(x)~(- \partial_{\mu}h_{C} \partial_{\nu}h_{D}
+ 2 \partial_{\mu}h_{C\rho\sigma} \partial_{\nu}h_{D}^{\rho\sigma}
+ 4 \partial^{\rho}h_{C\mu\sigma} \partial^{\sigma}h_{D\nu\rho}
\nonumber\\
+ 2 \partial_{\rho}h_{C\mu\nu} \partial^{\rho}h_{D}
- 4 \partial_{\rho}h_{C\mu\sigma} \partial^{\rho}h_{D\nu}^{\sigma} )(y)
\nonumber\\
+ 2 (\partial_{\rho}u_{A}^{\lambda}\partial_{\lambda}h_{B}^{\mu\nu})(x)~(\partial_{\lambda}h_{C} \partial_{\sigma}h_{D}
- 2 \partial_{\lambda}h_{C\mu\nu} \partial_{\sigma}h_{D}^{\mu\nu}
- 4 \partial^{\nu}h_{C\mu\lambda} \partial^{\mu}h_{D\nu\sigma}
\nonumber\\
- 2 \partial_{\mu}h_{C\sigma\lambda} \partial^{\mu}h_{D}
+ 4 \partial^{\mu}h_{C\lambda}^{\nu} \partial_{\mu}h_{D\nu\sigma} )(y)
\nonumber\\
+ (\partial_{\lambda}u_{A}^{\lambda}h_{B}^{\mu\nu})(x)~(- \partial_{\mu}h_{C} \partial_{\nu}h_{D}
+ 2 \partial_{\mu}h_{C\rho\sigma} \partial_{\nu}h_{D}^{\rho\sigma}
+ 4 \partial^{\sigma}h_{C\mu\rho} \partial^{\rho}h_{D\nu\sigma}
\nonumber\\
+ 2 \partial_{\rho}h_{C\mu\nu} \partial^{\rho}h_{D}
- 4 \partial_{\rho}h_{C\mu\sigma} \partial^{\rho}h_{D\nu}^{\sigma} )(y) ]
\eea
\bea
A_{h}^{\mu}(x,y) = 4~t^{ABCD}~[ (u_{A}^{\lambda}\partial_{\lambda}h_{B})(x)~(h_{C}^{\mu\nu} \partial_{\nu}h_{D}
- h_{C\rho\sigma} \partial^{\mu}h_{D}^{\rho\sigma})(y)
\nonumber\\
+  (u_{A}^{\lambda}\partial_{\lambda}h_{B\rho\sigma})(x)~(- 2 h_{C}^{\mu\nu} \partial_{\nu}h_{D}^{\rho\sigma}
- 4 h_{C\nu}^{\rho} \partial^{\sigma}h_{D}^{\mu\nu} 
- h_{C}^{\rho\sigma} \partial^{\mu}h_{D}
+ 4 h_{C\nu}^{\sigma} \partial^{\mu}h_{D}^{\nu\rho} )(y)
\nonumber\\
+ 2 (\partial_{\rho}u_{A}^{\lambda} h_{B}^{\rho\sigma})(x)~(2 h_{C}^{\mu\nu} \partial_{\nu}h_{D\lambda\sigma}
+ 2 h_{C\nu\lambda} \partial_{\sigma}h_{D}^{\mu\nu}
+ 2 h_{C\nu\sigma} \partial_{\lambda}h_{D}^{\mu\nu}
\nonumber\\
+ h_{C\nu\sigma} \partial^{\mu}h_{D}
- 2 h_{C\sigma}^{\nu} \partial^{\mu}h_{D\nu\lambda}
- 2 h_{C\lambda}^{\nu} \partial_{\mu}h_{D\nu\sigma} )(y)
\nonumber\\
+ 2 (\partial_{\rho}u_{A\sigma}h_{B}^{\rho\sigma})(x)~( - h_{C}^{\mu\nu} \partial_{\nu}h_{D}
+ h_{C\nu\lambda} \partial^{\mu}h_{D}^{\nu\lambda})(y)
\nonumber\\
+ (\partial_{\lambda}u_{A}^{\lambda}h_{B})(x)~( h_{C}^{\mu\nu} \partial_{\nu}h_{D}
- h_{C\nu\lambda} \partial^{\mu}h_{D}^{\nu\lambda})(y)
\nonumber\\
+ (\partial_{\lambda}u_{A}^{\lambda}h_{B}^{\rho\sigma})(x)~( - 2 h_{C}^{\mu\nu} \partial_{\nu}h_{D\rho\sigma}
- 4 h_{C\nu\sigma} \partial_{\rho}h_{D}^{\mu\nu}
- h_{C\rho\sigma} \partial^{\mu}h_{D}
+ 4 h_{C\nu\rho} \partial^{\mu}h_{D\sigma}^{\nu})(y) ]\quad
\eea
%\newpage

\bea
A_{gh}(x,y) = 8~t^{ABCD}~[ (u_{A}^{\mu}\partial_{\mu}h_{B\rho\sigma})(x)~
(\partial^{\rho}u_{C}^{\lambda} \partial_{\lambda}\tilde{u}_{D}^{\sigma}
- \partial_{\lambda}u_{C}^{\lambda} \partial^{\rho}\tilde{u}_{D}^{\sigma}
+ \partial^{\rho}u_{C}^{\lambda} \partial^{\sigma}\tilde{u}_{D\lambda})(y)
\nonumber\\
+ (\partial_{\rho}u_{A}^{\nu} h_{B}^{\rho\sigma})(x)~
(- \partial_{\sigma}u_{C\lambda} \partial^{\lambda}\tilde{u}_{D\nu}
- \partial_{\nu}u_{C\lambda} \partial^{\lambda}\tilde{u}_{D\sigma}
+ \partial_{\lambda}u_{C}^{\lambda} \partial_{\nu}\tilde{u}_{D\sigma}
\nonumber\\
+ \partial_{\lambda}u_{C}^{\lambda} \partial_{\sigma}\tilde{u}_{D\lambda}
- \partial_{\sigma}u_{C\lambda} \partial_{\nu}\tilde{u}_{D}^{\lambda}
+ \partial_{\nu}u_{C\lambda} \partial_{\sigma}\tilde{u}_{D\lambda})(y)
\nonumber\\
- (\partial_{\nu}u_{A}^{\nu} h_{B\rho\sigma})(x)~
(\partial^{\rho}u_{C}^{\lambda} \partial_{\lambda}\tilde{u}_{D}^{\sigma}
- \partial_{\lambda}u_{C}^{\lambda} \partial^{\rho}\tilde{u}_{D}^{\sigma}
+ \partial^{\rho}u_{C\lambda} \partial^{\sigma}\tilde{u}_{D}^{\lambda})(y)
\nonumber\\
+ {1 \over 2}~(\partial_{\nu}u_{A}^{\nu} h_{B})(x)~
(\partial_{\rho}u_{C}^{\rho} \partial_{\sigma}\tilde{u}_{D}^{\sigma}(y)
\nonumber\\
- (u_{A}^{\mu} \partial_{\mu}u_{B\nu})(x)
(\partial_{\rho}u_{C\sigma} \partial^{\nu}h_{D}^{\rho\sigma}(y) ]
\eea
\bea
A_{gh}^{\mu}(x,y) = 8~t^{ABCD}~[ (u_{A}^{\nu}\partial_{\nu}h_{B\rho\sigma})(x)~
(u_{C}^{\mu} \partial^{\rho}\tilde{u}_{D}^{\sigma}(y)
\nonumber\\
- (\partial_{\rho}u_{A\nu} h_{B}^{\rho\sigma})(x)~
(u_{C}^{\mu} \partial^{\nu}\tilde{u}_{D\sigma}
+ u_{C}^{\mu} \partial_{\sigma}\tilde{u}_{D}^{\nu})(y)
\nonumber\\
+ (\partial_{\nu}u_{A}^{\nu} h_{B\rho\sigma})(x)~(u_{C}^{\mu} \partial^{\rho}\tilde{u}_{D}^{\sigma})(y)
\nonumber\\
- (u_{A\sigma} \partial^{\sigma}u_{B}^{\rho})(x)~
(h_{C}^{\mu\nu} \partial_{\rho}\tilde{u}_{D\nu}
+ h_{C}^{\mu\nu} \partial_{\nu}\tilde{u}_{D\rho})(y)
\nonumber\\
+ (u_{A}^{\nu} \partial_{\nu}u_{B}^{\mu})(x)~
(\partial_{\rho}u_{C\sigma} h_{D}^{\rho\sigma}(y) ]
\eea
\be
t^{ABCD} \equiv t^{ABE} t^{CDE}
\ee
\newpage
The next step is to make an off-shell renormalization i.e. to consider the expressions:
\bea
T^{R}(T^{\mu}(x),T(y)) \equiv T(T^{\mu}(x),T(y)) - i \square D^{F}(x - y) A^{\mu}(y,x)
\nonumber\\
T^{R}(T(x), T^{\mu}(y)) \equiv T^{R}(T^{\mu}(y),T(x))
\nonumber\\
T^{R}(T(x),T(y)) \equiv T(T^{\mu}(x),T(y)).
\eea
For these new expresions we have:
\be
sT^{R}(T(x),T(y)) = \square D^{F}(x - y) [{\cal A}(x,y) + (x \leftrightarrow y) ]
\label{sTR}
\ee
where 
\be
{\cal A} = {\cal A}_{h} + {\cal A}_{gh}.
\ee
The explicit expressions are:
\bea
{\cal A}_{h}(x,y) = 2~t^{ABCD}~[ 2 (u_{A}^{\lambda}\partial_{\lambda}h_{B})(x)~
(\partial_{\mu}h_{C}^{\mu\nu} \partial_{\nu}h_{D}
+  h_{C}^{\mu\nu} \partial_{\mu}\partial_{\nu}h_{D}
- \partial_{\mu}h_{C\rho\sigma} \partial^{\mu}h_{D}^{\rho\sigma})(y)
\nonumber\\
+ (u_{A}^{\lambda}\partial_{\lambda}h_{B}^{\mu\nu})(x)~
( - 4 \partial_{\rho}h_{C}^{\rho\sigma} \partial_{\sigma}h_{D\mu\nu}
- 4 h_{C}^{\rho\sigma} \partial_{\rho}\partial_{\sigma}h_{D\mu\nu}
- 8 \partial_{\rho}h_{C\mu\sigma} \partial_{\nu}h_{D}^{\rho\sigma}
- 8 h_{C\mu\sigma} \partial_{\nu}\partial_{\rho}h_{D}^{\rho\sigma}
\nonumber\\
-  \partial_{\mu}h_{C} \partial_{\nu}h_{D}
+ 2 \partial_{\mu}h_{C\rho\sigma} \partial_{\nu}h_{D}^{\rho\sigma}
+ 4 \partial^{\rho}h_{C\mu\sigma} \partial^{\sigma}h_{D\nu\rho}
+ 4 \partial_{\rho}h_{C\mu}^{\sigma} \partial^{\rho}h_{D\nu\sigma})(y)
\nonumber\\
+ 2 (\partial_{\rho}u_{A}^{\lambda} h_{B}^{\rho\sigma})(x)~
( 4 \partial_{\mu}h_{C}^{\mu\nu} \partial_{\nu}h_{D\lambda\sigma}
+ 4 h_{C}^{\mu\nu} \partial_{\mu}\partial_{\nu}h_{D\lambda\sigma}
+ 4 \partial_{\mu}h_{C\nu\lambda} \partial_{\sigma}h_{D}^{\mu\nu}
+ 4 h_{C\nu\lambda} \partial_{\mu}\partial_{\sigma}h_{D}^{\mu\nu}
\nonumber\\
+ 4 \partial_{\mu}h_{C\nu\sigma} \partial_{\lambda}h_{D}^{\mu\nu}
+ 4 h_{C\nu\sigma} \partial_{\mu}\partial_{\lambda}h_{D}^{\mu\nu}
- 4 \partial_{\mu}h_{C\lambda}^{\nu} \partial^{\mu}h_{D\nu\sigma}
+ \partial_{\lambda}h_{C} \partial_{\sigma}h_{D}
\nonumber\\
- 2 \partial_{\lambda}h_{C\mu\nu} \partial_{\sigma}h_{D}^{\mu\nu}
- 4 \partial^{\nu}h_{C\lambda\mu} \partial^{\mu}h_{D\nu\sigma})(y)
\nonumber\\
+ 4 (\partial_{\rho}u_{A\sigma} h_{B}^{\rho\sigma})(x)~
( - \partial_{\mu}h_{C}^{\mu\nu} \partial_{\nu}h_{D}
- h_{C}^{\mu\nu} \partial_{\mu}\partial_{\nu}h_{D}
+ \partial_{\mu}h_{C\nu\lambda} \partial_{\mu}h_{D}^{\lambda\nu})(y)
\nonumber\\
+ 2 (\partial_{\lambda}u_{A}^{\lambda} h_{B})(x)~
( \partial_{\mu}h_{C}^{\mu\nu} \partial_{\nu}h_{D}
+ h_{C}^{\mu\nu} \partial_{\mu}\partial_{\nu}h_{D}
- \partial_{\mu}h_{C\rho\sigma} \partial^{\mu}h_{D}^{\rho\sigma})(y)
\nonumber\\
+ (\partial_{\lambda}u_{A}^{\lambda} h_{B}^{\mu\nu})(x)~
( - \partial_{\mu}h_{C} \partial_{\nu}h_{D}
+ 2 \partial_{\mu}h_{C\rho\sigma} \partial_{\nu}h_{D}^{\rho\sigma}
+ 4 \partial^{\sigma}h_{C\mu\rho} \partial^{\rho}h_{D\nu\sigma}
+ 4 \partial_{\rho}h_{C\mu\sigma} \partial^{\rho}h_{D\nu}^{\sigma}
\nonumber\\
- 4 \partial_{\rho}h_{C\mu\nu} \partial_{\sigma}h_{D}^{\rho\sigma}
- 4 h_{C}^{\rho\sigma} \partial_{\rho}\partial_{\sigma}h_{D\mu\nu}
- 4 \partial_{\rho}h_{C\nu\sigma} \partial_{\mu}h_{D}^{\rho\sigma}
- 8 h_{C\nu\sigma} \partial_{\mu}\partial_{\rho}h_{D}^{\rho\sigma})(y) ]
\eea
\bea
{\cal A}_{gh}(x,y) = 8~t^{ABCD}~[ (u_{A}^{\mu}\partial_{\mu}h_{B\rho\sigma})(x)~
(\partial_{\rho}u_{C}^{\lambda} \partial_{\lambda}\tilde{u}_{D}^{\sigma}
+ \partial^{\rho}u_{C}^{\lambda} \partial^{\sigma}\tilde{u}_{D\lambda}
+ u_{C}^{\nu} \partial_{\nu}\partial^{\rho}\tilde{u}_{D}^{\sigma})(y)
\nonumber\\
- (\partial_{\rho}u_{A}^{\nu} h_{B}^{\rho\sigma})(x)~
(\partial_{\sigma}u_{C\lambda} \partial^{\lambda}\tilde{u}_{D}^{\nu}
+ \partial_{\nu}u_{C\lambda} \partial^{\lambda}\tilde{u}_{D\sigma}
+ \partial_{\sigma}u_{C\lambda} \partial_{\nu}\tilde{u}_{D}^{\lambda}
\nonumber\\
+ \partial_{\nu}u_{C\lambda} \partial_{\sigma}\tilde{u}_{D\lambda}
+ u_{C}^{\mu} \partial_{\mu}\partial_{\nu}\tilde{u}_{D\nu}
+ u_{C}^{\mu} \partial_{\mu}\partial_{\sigma}\tilde{u}_{D\nu})(y)
\nonumber\\
+ (\partial_{\nu}u_{A}^{\nu} h_{B\rho\sigma})(x)~
(\partial^{\rho}u_{C}^{\lambda} \partial_{\lambda}\tilde{u}_{D}^{\sigma}
+ \partial^{\rho}u_{C}^{\lambda} \partial^{\sigma}\tilde{u}_{D\lambda}
+ u_{C}^{\mu} \partial_{\mu}\partial^{\rho}\tilde{u}_{D}^{\sigma})(y)
\nonumber\\
+ {1 \over 2}~(\partial_{\nu}u_{A}^{\nu} h_{B})(x)~
(\partial_{\rho}u_{C}^{\rho} \partial_{\sigma}\tilde{u}_{D}^{\sigma}(y)
\nonumber\\
- (u_{A\sigma} \partial^{\sigma}u_{B}^{\rho})(x)
(\partial_{\mu}h_{C}^{\mu\nu} \partial_{\rho}\tilde{u}_{D\nu}
+ \partial_{\mu}h_{C}^{\mu\nu}  \partial_{\nu}\tilde{u}_{D\rho}
+ h_{A}^{\mu\nu} \partial_{\mu}\partial_{\nu}\tilde{u}_{D\rho})(y) ]
\eea
%\newpage
In the on-shell limit we have
\be
\square D^{F}(x - y) \rightarrow \delta(x - y)
\ee
so the formula (\ref{sTR}) becomes
\be
sT^{R}(T(x),T(y)) = \delta(x - y) {\cal A}(x)
\label{sTR-shell}
\ee
where
\be
{\cal A}(x) \equiv 2 {\cal A}(x,x).
\ee
Then remaining anomaly
$
{\cal A}
$
can be eliminated {\it iff} it can be written under the form 
\be
{\cal A} = \bar{s}N = d_{Q}N + i \partial_{\mu}N^{\mu}
\label{A-NN}
\ee
i.e. it is a coboundary. We consider only the first contribution and note that it has the structure:
\be
{\cal A}_{h} = u_{A}^{\mu} H_{\mu}^{A} + \partial^{\nu}u_{A}^{\mu} H_{\mu\nu}^{A}
\ee
where
$
H_{\mu}^{A}
$
and
$
H_{\mu\nu}^{A}
$
are expressions tri-linear in
$
h_{B}^{\alpha\beta}.
$
From (\ref{A-NN}) it follows that the terms of the first kind (i.e. without derivatives on 
$
u_{A}^{\mu}
$)
can appear only from
$
N^{\mu}
$
more precisely we must have:
\be
N^{\mu} = u_{A\nu} t^{\mu\nu}_{A} + \cdots 
\ee
where
$
t^{\mu\nu}_{A}
$
are expressions tri-linear in
$
h_{B}^{\alpha\beta}.
$

Then it easily follows that we must have:
\be
H_{\mu\nu}^{A} = \partial^{\nu}t_{A\nu\mu}.
\label{Ht}
\ee

It is sufficient to consider the terms with two factors $h_{B}$ and one factor
$
h_{B}^{\alpha\beta}
$
from
$
H_{\mu\nu}^{A}
$
and write the generic anszatz for the corresponding sector of 
$
t^{\mu\nu}_{A}
$:
\bea
t^{A(1)}_{\alpha\lambda} = t_{1}^{ABCD} h_{B\alpha\lambda} \partial^{\beta}h_{C} \partial_{\beta}h_{D} \qquad 
%\nonumber\\
t^{A(2)}_{\alpha\lambda} = t_{2}^{ABCD} h_{B\alpha\beta} \partial_{\lambda}h_{C} \partial^{\beta}h_{D}
\nonumber\\
t^{A(3)}_{\alpha\lambda} = t_{3}^{ABCD} h_{B\lambda\beta} \partial_{\alpha}h_{C} \partial^{\beta}h_{D} \qquad 
%\nonumber\\
t^{A(4)}_{\alpha\lambda} = t_{4}^{ABCD} \eta_{\alpha\lambda} h_{B}^{\beta\gamma} \partial_{\alpha}h_{C} \partial_{\gamma}h_{D}
\nonumber\\
t^{A(5)}_{\alpha\lambda} = t_{5}^{ABCD} \partial_{\alpha}h_{B\lambda\beta} h_{C} \partial^{\beta}h_{D} \qquad
%\nonumber\\
t^{A(6)}_{\alpha\lambda} = t_{6}^{ABCD} \partial_{\lambda}h_{B\alpha\beta} h_{C} \partial^{\beta}h_{D}
\nonumber\\
t^{A(7)}_{\alpha\lambda} = t_{7}^{ABCD} \partial_{\beta}h_{B\alpha\lambda} h_{C} \partial^{\beta}h_{D} \qquad
%\nonumber\\
t^{A(8)}_{\alpha\lambda} = t_{8}^{ABCD} \partial^{\beta}h_{B\alpha\beta} h_{C} \partial_{\lambda}h_{D}
\nonumber\\
t^{A(9)}_{\alpha\lambda} = t_{9}^{ABCD} \partial^{\beta}h_{B\lambda\beta} h_{C} \partial_{\alpha}h_{D} \qquad
%\nonumber\\
t^{A(10)}_{\alpha\lambda} = t_{10}^{ABCD} \eta_{\alpha\lambda} \partial^{\gamma}h_{B\beta\gamma} h_{C} \partial^{\beta}h_{D}
\nonumber\\
t^{A(11)}_{\alpha\lambda} = t_{11}^{ABCD} h_{B\alpha\beta} h_{C} \partial_{\lambda}\partial^{\beta}h_{D} \qquad
%\nonumber\\
t^{A(12)}_{\alpha\lambda} = t_{12}^{ABCD} h_{B\lambda\beta} h_{C} \partial_{\alpha}\partial^{\beta}h_{D}
\nonumber\\
t^{A(13)}_{\alpha\lambda} = t_{13}^{ABCD} \eta_{\alpha\lambda} h_{B\beta\gamma} h_{C} \partial^{\beta}\partial^{\gamma}h_{D} \qquad
%\nonumber\\
t^{A(14)}_{\alpha\lambda} = t_{14}^{ABCD} \partial_{\alpha}\partial^{\beta}h_{B\lambda\beta} h_{C} h_{D}
\nonumber\\
t^{A(15)}_{\alpha\lambda} = t_{15}^{ABCD} \partial_{\lambda}\partial^{\beta}h_{B\alpha\beta} h_{C} h_{D} \qquad
%\nonumber\\
t^{A(16)}_{\alpha\lambda} = t_{16}^{ABCD} \eta_{\alpha\lambda} \partial_{\beta}\partial_{\gamma}h_{B}^{\beta\gamma} h_{C} h_{D}
\eea
where
\be
t_{j}^{ABCD} = C \leftrightarrow D, \quad j = 1, 4, 14, 15, 16.
\ee

We insert everything in the relation (\ref{Ht}) and obtain a linear system. An easy consequence of this system is
\be
t^{ABCD} = A \leftrightarrow C.
\ee
Now the proof goes as in \cite{BDGH}: we define the 
$
R \times R
$
matrices
\be
(T_{A})_{BC} \equiv t^{ABC}
\ee
and the previous relation can be written as:
\be
T_{A} T_{B} = T_{B} T_{A}
\ee
Because the matrices
$
T_{A}
$
are real, symmetric and commute they can be diagonalized simultaneously i.e in a convenient base we have:
\be
t^{ABC} = \lambda_{AB} \delta_{BC}.
\ee
From here it follows that only the expressions
$
t^{AAA}
$
can be non-zero. It means that there are no cross terms of interactions between two distinct gravitons. This is our main result.
%\newpage
The case of massive gravitions \cite{massive} leads to the same no-interaction theorem. Indeed, in the massive case the terms 
of top canonical dimension
$
\omega = 5
$
from all expresions are the same as in the massless case.

For completness we finish the analysis in the case 
$
R = 1
$
i.e. a single gravition, so we do not need the indices
$
A, B, \dots.
$
It can be proved rather easily that the ghost part of the anomaly
$
{\cal A}_{gh}
$
is a total derivative. So the relation (\ref{A-NN}) must be imposed only on
$
{\cal A}_{h}.
$
First, we rewrite it in the form
\be
{\cal A}_{h} = u_{\mu} {\cal A}_{h}^{\mu} + i~\partial_{\mu}N_{h}^{\mu}
\ee
with 
$
{\cal A}_{h}^{\mu} \sim h h h.
$
Then the following result can be proved by some computations (see also \cite{Sc2}):
\begin{thm}
The following formula is true
\be
{\cal A}_{h} = d_{Q}N + \partial_{\mu}N^{\mu}
\ee
where:
\bea
N = 4 i (2 h^{\mu\nu} h^{\rho\sigma} \partial_{\rho}h_{\mu\nu} \partial_{\sigma}h
- h^{\mu\nu} h^{\rho\sigma} \partial_{\lambda}h_{\mu\nu} \partial^{\lambda}h_{\rho\sigma}
- 4 h^{\mu\nu} h_{\nu\rho} \partial^{\lambda}h^{\rho\sigma} \partial_{\sigma}h_{\mu\lambda}
\nonumber\\
- 4 h^{\mu\nu} h^{\rho\sigma} \partial_{\mu}h_{\rho\lambda} {\partial_{\nu}h_{\sigma}}^{\cdot\lambda}
+ 4 h^{\mu\nu} h_{\nu\rho} \partial_{\lambda}h_{\mu\sigma} \partial^{\lambda}h^{\rho\sigma}
+ 2 h^{\mu\nu} h^{\rho\sigma} \partial_{\lambda}h_{\mu\rho} \partial^{\lambda}h_{\nu\sigma}
- 2 h^{\mu\rho} {h_{\nu}}_{\dot\rho} \partial_{\lambda}h_{\mu\nu} \partial^{\lambda}h).
\label{second-order}
\eea
\end{thm}
The idea of the proof is similar to previous one. We compute the expressions of the type
$
d_{Q}N_{j}
$
for various Wick monomials
$
N_{j} \sim h h h h
$
and of canonical dimension
$
\omega = 6.
$
Next, ``by partial integration" we exhibit them in the form 
\be
d_{Q}N_{j} = \partial_{\mu}M_{j}^{\mu} + u_{\mu}~N_{j}^{\mu}.
\ee
If we succeed to fix the coefficients of these monomials such that
\be
\sum a_{j} N_{j}^{\mu} = {\cal A}_{h}^{\mu}
\ee
then the proof is finished. The monomials from the statement do the job. We can eliminate the anomaly
$
{\cal A}
$
by obvious redefinitions of the chronological products. We remark that the redefinition $N$ of
$
T(T(x),T(y))
$
does not contain ghost terms. This seems to be the main advantage of the choice (\ref{T}).
\newpage
\section{Conclusions}
We have derived the no-interaction theorem for the multi-graviton system in the framework of the causal formalism of 
perturbative quantum field theory. However, interaction between two distinct species of gravitions might be possible
mediated by matter fields, in higher orders of perturbation theory. This is a subject of further invesigation.

\end{document}